\documentclass{aastex631}
\usepackage[utf8]{inputenc} % Allows UTF-8 input characters
\usepackage[T1]{fontenc}    % Optimizes font encoding for accents
\usepackage{booktabs}       % Creates beautiful, clean table lines
\usepackage{pgfplotstable}
\usepackage{makecell}

\begin{document}

\title{The shell around VY Canis Majoris in SPHEREx is likely an artifact}

\author[0000-0003-2758-159X]{Viraj Karambelkar}
\email{vk2588@columbia.edu}
\altaffiliation{NASA Hubble Fellow}
\affiliation{Columbia University, 538 West 120th Street 704, MC 5255, New York, NY 10027}

\author{Yashvi Sharma}
\affiliation{Caltech Optical Observatories, California Institute of Technology, Pasadena, CA 91125, USA}

\author[0000-0002-8989-0542]{Kishalay De}
\affiliation{Columbia University, 538 West 120th Street 704, MC 5255, New York, NY 10027}
\affiliation{Center for Computational Astrophysics, Flatiron Research Institute, 162, 5th Ave, New York, NY 10010}

% \author{Daniel Weatherill}
% \affiliation{Caltech Optical Observatories, California Institute of Technology, Pasadena, CA 91125, USA}
\begin{abstract}
    Recently, the presence of a large shell around the bright hypergiant star VY Canis Majoris was reported in images from the SPHEREx mission. We report similar shells around other bright stars at similar locations in SPHEREx Band 1 images, suggesting that this shell is not astrophysical but instead a detector artifact. 
\end{abstract}

\section*{The shell around VY CMa} 
\label{sec:intro}
Recently, the presence of a large shell was reported around the bright hypergiant star VY Canis Majoris (VY CMa) \citep{Banerjee2026} in images from the SPHEREx mission \citep{Bock2026}. The shell is clearly visible in two SPHEREx Band 1 images, in both of which the position of the star on the detector corresponds to a wavelength of $\approx7827$\,A. The shell has a radius of about 40 pixels, and extends over pixels corresponding to wavelengths between 7700 and 7900\,A. This shell is not visible in deep ground-based imaging or in spectroscopic observations \citep{Banerjee2026}. 

\begin{figure}
    \centering
    \includegraphics[width=\textwidth]{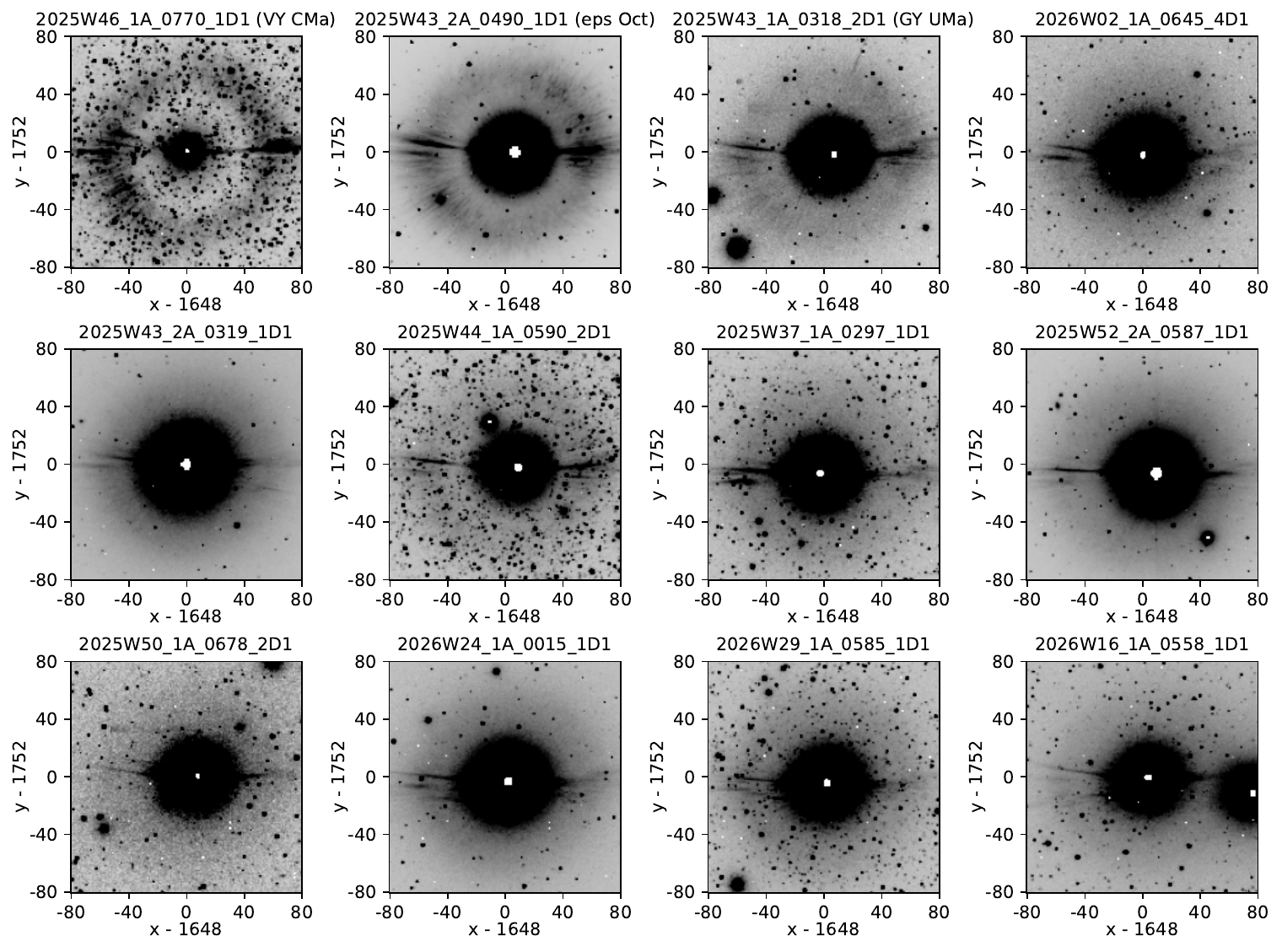}
    \caption{Cutouts of SPHEREx Band 1 images that have bright stars located close to pixel coordinates x=1648 and y=1752. The stars VY CMa, eps Oct, and GY UMa which show a clear shell are shown in the first three panels. The contrast of the shell is lower in other stars, but they all clearly show some bright spikes similar to VY CMa and eps Oct. All cutouts are plotted with a stretch of $+/-3\sigma$. SPHEREx image identifiers are listed for each cutout.}
    \label{fig:xx}
\end{figure}

Motivated by the similarity of the spikey-structure of this shell (Figure\,\ref{fig:xx}) with detector artifacts from charge-blooming and bad-pixels around bright sources in SPHEREx \footnote{\url{https://irsa.ipac.caltech.edu/data/SPHEREx/docs/SPHEREx_Expsupp_QR_v2.0.pdf}} (Fig. 2 in \citealt{Fazar2026}), we searched through other Band 1 SPHEREx images for similar shells to investigate a possible artifact origin. We searched through $\approx$230000 publicly available Band 1 SPHEREx QR2 images \citep{spherex_qr2}, and flagged 110 images that have a bright star close to the exact location of VY CMa. Specifically, we required at least one pixel with counts exceeding 2000 within a 20-pixel box around x=1648 and y=1752, which is the central position of VY CMa in the two images where the shell is visible. We examine these 110 images, and identify several instances of bright stars surrounded by structures resembling that seen around VY CMa. Some examples are shown in Figure\,\ref{fig:xx}. In particular, the stars eps Oct and GY Uma clearly show the exact same spikey-shell structure as VY CMa. The contrast of shells around other stars is lower than these two, however, they all clearly show the bright spikes that are seen around VY CMa. 

The presence of similar structures in multiple images at similar positions on the detector suggests that the shell around VY CMa is likely a detector artifact and not astrophysical. This has now been recognized to be an artifact by \citet{Banerjee2026} too, to whom the results by \citet{Fazar2026} (published only on 26 August 2026) were not available at the time of manuscript submission and processing. The arxiv submission of \citet{Banerjee2026} has now been withdrawn (see arxiv version 2 for details). 

This artifact appears to be specific to this region on the Band 1 detector - generally, we did not find similar artifacts (full shells or just spikes) around other bright stars located elsewhere on the detector. We also find several cases where a bright star in this region does not show any spikes or shell structures - so this artifact appears to require conditions beyond just the presence of a bright star in this vicinity.

\section*{Acknowledgements}
We thank Dan Weatherill and D.P.K. Banerjee for helpful discussions. This publication makes use of data products from the Spectro-Photometer for the History of the Universe, Epoch of Reionization and Ices Explorer (SPHEREx), which is a joint project of the Jet Propulsion Laboratory and the California Institute of Technology, and is funded by the National Aeronautics and Space Administration. We are also grateful to the Center for Computational Astrophysics, Flatiron Institute for providing storage and computational resources used in this study.
\bibliography{myreferences}{}
\bibliographystyle{aasjournal}

\end{document}